\documentstyle[prl,aps]{revtex}
\begin{document}

\vskip 4mm

\centerline{\large \bf Temperature Effect on the Operation}
\centerline{\large \bf of Elementary Quantum-Dot Spin Gates}
\centerline{\large \bf by the Example of the NOT-AND Gate}

\vskip 2mm

\centerline{Arcadiy V.Krasheninnikov and Roman A.Koltsov}

\vskip 2mm

\centerline{\it Moscow State Engineering Physics Institute
(Technical University),}
\centerline{\it Moscow, 115409, Russia}

\vskip 4mm

\begin{quotation}

The effect of temperature on
the operation of the elementary quantum-dot spin gates for single-electron
computing is studied theoretically within the framework of
the Hubbard model by the example of the NOT-AND gate.
The calculated values of the uniform external magnetic field necessary
to realize the whole truth table
proved to be unreasonably high for the implementation of
the NOT-AND logical function even at the liquid helium temperature.
This result appears to be common to all spin gates.
Thus, finite temperatures seem to be a serious obstacle
for the practical realization of ground state calculations
in quantum dot spin gates.

\end{quotation}

\vskip 6mm

{\bf 1. Introduction}

\vskip 2mm

The problem of miniaturization of electronic computer components
attracts much attention nowadays. The Si-based MOSFET technology is
expected to enter a critical period in the beginning of the next
millennium, since the minimum feature size of the computer basic unit,
the transistor, will be about 0.1 micron that time if
the empirical trend called as ``Moore Law'' \cite{Moore} will remain correct
in the near future. Further reduction
in size seems to be inefficient \cite{Luth}. Indeed, if even technological
problems such as uncontrolled variation in transistor characteristics from
specimen to specimen due to nonuniform distribution of doping atoms, Joule
resistance heating, electromigration of atoms
resulting from the extremely high current densities, {\it etc.} will be
circumvented, there appear the fundamental limitations on
the transistor operation related with
the laws of quantum mechanics. Obviously, the classical picture is not correct
on the nanometer scale, and a device must function based
on quantum effects such as energy quantization and tunneling rather than
in spite of them \cite{techn}.  Thus, the future of computing is closely
related with the development of intrinsically quantum
nanometer-scale replacements for the bulk-effect semiconductor transistors.

Although the operating principles of even individual quantum devices and
the interconnection problem are challenging tasks for scientists,
advantages of nanocomputers such as high operation speed and low power consumption
have been an enormous stimulus for the development of various approaches
to the computations on the nanoscale.

Among them the theory of quantum computation is an extremely exciting and rapidly
growing field of investigation \cite{PW}. This concept
originating from the early works of Richard Feynman \cite{Feynman}
and David Deutsch \cite{Deutsch} in 1980s is based on
the superposition, interference, entanglement and other fundamental
properties \cite{Bennett} of a quantum system. Quantum computer
requires quantum logic: it deals with the quantum bits or qubits,
for short, i.e.
with an arbitrary superposition of pure classical bits, 0 and 1.
Unlike its classical counterpart, quantum computers use ``quantumness''
at the information-theoretical level. Quantum computations are
reversible, and all states in the superposition of qubits are processed
simultaneously, so the calculation may be speeded up exponentially
\cite{Plenio}. The Shor's \cite{Shor} discovery
of quantum algorithms for the integer factorization and for the discrete
logarithm, that run exponentially faster \cite{Grover} than
the best known classical algorithms
and the experimental demonstrations \cite{Turchette}, \cite{Monroe}
of a single elementary unit of a quantum computer gave rise to a
dramatic growth in the number of publications on quantum computations.
At the same time, practical implementation of a quantum
computer and, in general, feasibility of quantum computations
in a physical system casts serious doubts
\cite{Plenio},  \cite{Landauer1},  \cite{Landauer2}.
The principal issue is the ratio of quantum computation speed
to the decoherence rate \cite{Monroe_PT}. There are fundamental limitations
\cite{Plenio}
related to the particular implementation of a quantum computer
(e.g., a rate at which lasers may drive atomic transitions of a given lifetime,
{\it etc.}) that inevitably decrease computational speed. So
the decoherence, inherent for any real physical system,
is believed never to be reduced to the point
where more than a few consecutive quantum computational steps can be
made \cite{Landauer1}.

Other approaches such as mechanical \cite{mech}, chemical \cite{Adleman1},
\cite{Lipton}  and electronic nanocomputing
\cite{Lent}, \cite{Bandyo}, \cite{Korotkov}
employ classical Boolen logic.
Upon development of these concepts many aspects of conventional
computation realized in semiconductor microcomputers are taken into account.
(Recall that quantum computational algorithms are shown to be more
efficient than classical ones for the two problems only \cite{Shor}, \cite{Grover}).
But the basic units of nanocomputers differ fundamentally  from usual
microtransistors, being about several nanometers in size.

The theoretical possibility for building mechanical nanocomputers
was de\-monstrated in \cite{mech,Hall}. Such computers, if
constructed, would operate like a vastly scaling
down programmable version of the mechanical calculators appeared in the
1940s. The data processing is performed with the help of moving
molecular-scale rods and rotating molecular-scale wheels assembled
by the mechanical positioning of atoms and molecules using STM.
The fabrication of such devices is tedious task demanding
the perfection and development of STM-technology. Besides,
other practical issues must be addressed, such as how to power and program
nanomachinery \cite{techn}.

The operation of a chemical nanocomputer is based on the storage of
data in chemical (i.e. molecular) structures. Information processing is
performed by making and breaking chemical bonds. The existing
biochemical variants of such computers are humans and animals nervous systems.
Unfortunately, the artificial fabrication of this category
of nanocomputers seems to be far beyond the experimental realization, since
the mechanisms for animal brains and nervous systems are poorly understood.
Nevertheless, a complex graph theory problem has been solved
recently on sequences of DNA's molecular subunits \cite{Adleman1}. This
approach being a giant leap towards the creation of a biochemical computer
may be applied not only to the solution of combinatorial problems, but
to a much wider class of digital computations \cite{Lipton}. Note that
despite a number of remaining obstacles such as efficient input and output,
error correction {\it etc.}, only this type of nanocomputers among mentioned above
is demonstrated for an actual calculation.

The information processing in electronic nanocomputers is related with
the storage and movement of electrons. There exist various propositions
to use a quantum mechanical nanosystem for calculations in such a way.

For example, Lent {\it et al.} \cite{Lent} suggested  constructing a two-state
cell made of five quantum dots having two electrons. These electrons can exist
inside the cell in two equally probable states, so the polarization of a
cell represents a binary zero or one. The polarization of an individual cell
may be fixed by applying an appropriate voltage on an external probe  or
due to effect of another cell. Rows of closely arranged quantum dot cells
can carry signals (i.e. change polarization), or realize various logical
functions (e.g. NOT, AND, OR etc.).

Korotkov {\it et al.} \cite{Korotkov} proposed for computing wire-like structures  composed of quantum
dots placed in a global electric field. Each local element is a row of quantum
dots. A neighboring element is a similar row of quantum
dots perpendicular to the first. A signal is propagated through the system
by the formation of
electron-hole pairs in each row of quantum dots. The electric field
allows the local polarization of the electron-hole pair in one element
which in turn induces the formation and polarization of a pair in the
next element. Thus, chains of elements could be linked together to
implement elementary logical functions.

The concept of computing by measuring spins of individual electrons on
quantum dots
was originally introduced by Bandyopadhay {\it et al.} \cite{Bandyo}.
In the context of this approach a computer is supposed to consist of
spatially arranged arrays of tunnel-coupled quantum dots on a substrate.
Each dot has single size-quantized energy level. The total number of electrons
in the system, controlled by adjusting the voltage of substrate \cite{Meurer}
is approximately equal to the number of dots, thus there is one electron
per dot on the average. Bits of information are spins of individual
electrons on a given quantum dot, e.g. the ``up'' (``down'') direction of
electron spin corresponds to the logical 1 (0). The temperature of the
system is supposed to be equal to zero. The information is
stored in the spin  configuration of the system, i.e.
it is conditioned by a set of quantum-mechanical ground state average values
of the spin projection operator $\hat{S}_{zi}$ on the $i$-th dot in the array.

Although it is not obvious {\it apriori}, it is believed that operation of
the whole ``spin computer'' may be described considering operation of
separated spin gates, by analogy with today's computers. ``Gate'' is an
elementary unit of a computer capable to perform simplest logical functions.
The example of spin gates is the ``NOT-AND'' gate, see Fig.1.
Each gate has input and output dots. The number of such dots in the gate
depends on its logical function. ``NOT-AND'' gate has two input (A, B)
and one output dot (Y). The former serve for writing the input
signals to the gate by the action of an external agent (e.g. local magnetic
field generated by a magnetic tip of STM). As a result of external influence
on input dots, the new ground state of the system is different from the
initial one, so the spin configuration represents the result of a
desired computational operation. Upon such a process the electron tunneling
between adjacent dots and Coulomb repulsion of electrons provide the
propagation of the ``signal'' from dot to dot. The result of calculation may be
read off from output dots by means of, e.g., magnetic tip since the
tunneling current depends on the mutual orientation of the magnetizations
of the dot and tip. The correspondence between magnetizations of output and
input dots is uniquely determined by the logical truth table of a particular
gate. The logical truth table of NOT-AND gate is shown in Fig.1.

The operation of spin gates was studied theoretically in \cite{Molotkov1},
within the spin-1/2 Heisenberg model by means of the exact diagonalization
technique. It was shown that for a number of the simplest logical gates, such
as NOT, AND, NAND, OR, NOR, NXOR and half-adder, the entire truth table can
be obtained by the appropriate choice of values of the local magnetic fields
on the input dots.
Since the logical variables 0 and 1 were
associated with ground state averages of the spin projection operator $\hat{S}_{zi}$
on i-th quantum dot, the threshold value $S_t$, ($0<S_t<1$) of the projection of
electron spin on the $z$ axis was introduced \cite{Molotkov1}. The case
$\langle\hat{S}_{zi}\rangle \ge S_t/2$ was supposed to correspond to
logical one, and $\langle\hat{S}_{zi}\rangle \le -S_t/2$ to logical zero.
It should be noted that in \cite{Molotkov1} this threshold value was chosen
to be sufficiently low, usually $S_t=0.05 \div 0.1$ (The entire truth
table could not be realized otherwise).
Such small values of $S_t$ result in a high probability of error upon
measurement (this problem is discussed at length in \cite{Openov2})
due to a quantum-mechanical nature of electron spin. The arising
necessity of ``redundancy'', i.e. the need for repeating the calculation
process several times or for additional logic gates that work
simultaneously will have deleterious effect on the operation
of ``spin computers''.  Besides, the physical truth tables, calculated
in \cite{Molotkov1} were not symmetrical with respect to the
input signal. E.g., the $011$ and $101$
rows of the logical truth table of the NOT-AND gate are never
realized at equal absolute values of local magnetic fields on
the gate input dots \cite{comp1}.

A new approach to the implementation of certain logical functions
in spin gates was proposed in \cite{comp1}. It was shown that
placing a gate in a {\it uniform} external magnetic field $H_z$
allows one to remove the lack of symmetry of the physical truth table
and enlarge the
absolute magnitudes of the average spins at the output dots.
It also was demonstrated \cite{Molotkov2} that the
introduction of ``ferromagnetic chains'' into the gate structure
and application of {\it local} constant magnetic fields acting on particular
dots can substantially improve the physical truth table of gates.
Thus, with the elimination of the ``small spin problem'' the
incorporation of gates into a large circuit may be possible.

Unfortunately, the operation of a spin gate computer has other
fundamental limitations. One of them is related with the ground
state computing concept itself. This concept underlies not only
spin gates computers, but many other semiconductor quantum-dots-based
logic devices \cite{Nomoto}.
Since the computation is related with the relaxation of the system
into a new ground state under the influence of the external source,
the speed of the operation is determined by
the dissipative coupling between the system and the environment.
The rate, at which the dissipation in bulk semiconductors occurs, is
conditioned predominantly  by the emission of a longitudinal optical phonon.
This rate is equal to $10^{12}$ s$^{-1}$. At the same time,
such a process is prohibited  in quantum dots
because of the discrete nature of the energy
spectrum, unless the level separation equals to the
longitudinal optical phonon energy. As was shown in \cite{Nomoto},
the longitudinal-acoustic phonon emission is the main
contributor to the relaxation process in quantum dots, since the acoustic
phonons have continuous spectra from zero up to a certain limiting energy.
The relaxation rate was predicted to be $(10^6 - 10^9)$ s$^{-1}$
\cite{Nomoto}. Obviously, such an operation speed is too low for
consideration
of semiconductor spin gates as contenders for a new generation
of logic devices. Note, however, that other relaxation mechanisms
(e.g. interface-phonon scattering) are possible giving rise
to the increase in the device functioning speed.

But even if the above limitations
will be circumvented, there remains the negative effect of finite
temperatures on the operation of a { \it real} spin gate computer.

If the temperature $T \ne 0$, quantum-mechanical ground state average values
of spin projection operators must be replaced with the thermodynamical
averages.
It is clear {\it apriori}, that the temperature
will decrease the values of $\hat{S}_{zi}$, resulting in the enhancement
of the error probability and in the ``shrinkage'' of the physical truth
table. But one may hope that increasing the external stabilizing
field may provide the operation of spin gates at the temperatures
that may be achieved in the reality, e.g. the liquid nitrogen temperature,
or at least, helium temperature.
It is the purpose of the paper to elucidate this issue and to establish
the quantitative relations between the temperature and other
parameters of the system, at which logic gates could be working.

\vskip 6mm

{ \bf 2. Theoretical Model for the Spin Gates}

Provided that each dot has a single size-quantized level, interacting
electrons on dots may be described by a suitably parametrized Hubbard
model \cite{Hubb} taking into account the tunneling and the intradot Coulomb repulsion
of electrons. We do not consider the interaction of electrons occupying
different dots, since its energy is much smaller than the intradot repulsion
(see also \cite{Openov2}.
The relevant Hamiltonian has the form:
\begin{equation}
\label{trivial}
H=-t\sum_{<ij>,\sigma } (a^{+}_{i\sigma } a_{j\sigma } + H.c.) -
\mu_B \sum_{i,\sigma } n_{i\sigma } H_i {\bf sign} \sigma+
U \sum_{i} n_{i \uparrow}  n_{i \downarrow}  ,
\end{equation}
where $a_{i\sigma}$ ($a^+_{i\sigma}$)  is the  operator of annihilation
(creation) of electron  on the
$i$-th dot  with spin projection $\sigma$ = +1 or -1  on the  $z$-axis,
$n_{i\sigma}=a^+_{i\sigma}a_{i\sigma}$ is the
electron  number  operator,  $t$  is  the  matrix element for hopping of
electrons between quantum  dots, $U$ is  the intradot Coulomb  repulsion
energy, $H_i$ is the local magnetic field (along z-axis) at the $i$-th dot,
$\mu_B$ is the Bohr magneton, $<ij>$ means the summation over nearest the
neighbor dots. In what follows, we set $\mu_B=1$.

It is supposed that the total number of electrons $N=\sum_{i,\sigma} n_{i\sigma}$
coincides with the number of dots in  the gate and the ratio $U/t$ is
large enough to provide strong antiferromagnetic correlations in the gate.
These correlations  result in  the  switching  of  electron  spins in the gate
after the local fields  $H_i$ have acted on the input dots.

In order to determine the physical truth
table, i.e. the range  of control
signals (local magnetic fields at the input dots) for which the logical truth
table of a particular gate is realized, it is necessary to calculate
the ground state wave function of the Hamiltonian (1) if $T = 0$ and
in general case all the wave functions if $T \ne 0$. For the
relatively small number of dots in  the gate this can be done
numerically by an exact diagonalization method \cite{diag}.
At final temperatures the quantum-mechanical ground state average value
of spin projection operator $\hat{S}_{zi}$ on the $i$-th dot in the array
is determined by the well-known formula
\begin{equation}
\langle \hat{S}_{zi} \rangle = \frac{\sum_{k} \langle k|\hat{S}_{zi}
\exp(-H/T)|k \rangle}
{\sum_{k} \langle k|\exp(-H/T)|k \rangle}
\end{equation}
where $|k\rangle,~~k=0,1,2...$ is the {\it k}-th  eigen function of the
Hamiltonian (1).

\vskip 6mm
{3. \bf Constructing the Physical Truth Table}

The procedure of the physical truth
table construction may be illustrated by the example of the NOT-AND gate.
Suppose that $T = 0$.
Having the ground state wave function obtained, one can determine
the resulting  values of  electron spin  on input (
$\langle S_A \rangle $, $\langle S_B \rangle $ ) and output ( $\langle S_Y \rangle $ )
dots after  the local fields have acted. If, e.g.,
$\langle S_A \rangle > S_t/2$, $\langle S_B \rangle > S_t/2$, and
$\langle S_Y \rangle < -S_t/2$  at given  values of  $H_A$ and  $H_B$, then the
first row (110)[ABY] of  the logical truth table  of the NOT-AND gate
is realized, and we mark  the point ($H_A,H_B$) in the $H_A-H_B$  plane
by the  symbol ``+'' (see Fig.2) .  If the  spin configuration  corresponds to one of
the other three rows  of the logical truth  table (011, 101, or  001),
then the  point ($H_A,H_B$)  is  marked by other symbol (since the domains
are well spatially separated for the NOT-AND gate we marked all the domains
by the same symbol). The blank space in the $H_A-H_B$ plane means that none of the rows of  the
logical truth table is realized at given values of $H_A$ and $H_B$.
The procedure is absolutely the same if $T \ne 0$.

\vskip 6mm

{\bf 4. Results of Calculations and Discussion}

The calculated physical truth tables of the NOT-AND gate at various
temperatures are shown in Fig.3 for $S_t=0.5$. We deliberately chose
the value of $S_t$ being not very large. Although the decrease in
$S_t$ gives rise to the enhancement in the probability to read off
the incorrect result of calculation, it is instructive to show the effect
of temperature on the physical truth table for such values of  $S_t$
when the range of temperatures at which the gate still works is not
too small (the quantitative relation between the threshold values
of $S_t$ and maximal operating temperature will be established below).

The parameters of Hamiltonian (1) and regions of the input
signals $H_A$ and $H_B$ are the same as in \cite{comp1}: $U/t=20,
H_z/t=0.05$. Notice that
the ratio $U/t$ is large enough to provide the antiferromagnetic
correlations in the system. The value of a uniform external
magnetic field $H_z$ is close to the optimal one, i.e. to the
field at which the projection of electron spin on the output dot
on the $z$ axis reaches its maximum \cite{comp1}. As follows from Fig.3,
the increase in temperature results in the shrinkage of domains in
the truth table. Different domains are differently
sensitive to the temperature: the domains where the
value of the projection of electron spin at $T=0$ is the
result of interplay between stabilizing external field and
``counteraction'' of antiparallel spin alignment on the input dots
are more sensitive to the increase in temperature.

In order to describe the temperature effect on the operation of
the NOT-AND gate more comprehensively, we plotted in Fig.4
the dependence of {\it modulus} of $\langle\hat{S}_{z}\rangle$ on the
output dot for the four
characteristic points (see Fig.2) in the physical truth table at
various values of $H_z$.
Actually, we present data for three points only, since points ``2'' and ``4''
are absolutely equivalent. Although our choice is rather
arbitrary, it is conditioned by the demand for the symmetry of
$H_A, H_B$ values. Besides, $H_A/t, H_B/t$ should not be very large.
Solid line corresponds to the point ``1'', dashed line stands for
the point ``2'', long dashed line for ``3''.

Let us introduce the critical temperature $T^*$ as a maximal temperature
at which the logical function of the NOT-AND gate can be implemented
(at fixed values of $H_A, H_B$ and $H_z$).  The procedure to determine
$T^*$ is as follows (see Fig.4b for the sake of definiteness). Having
fixed the value of $S_t$ (in our case $S_t=0.5$), we draw a horizontal
line in the plot $ \langle\hat{S}_{z}\rangle$ vs. $T$. We remind that the
line $ \langle\hat{S}_{z}\rangle=0.25$ corresponds to the value $S_t=0.5$.
The first intersection of the horizontal line with curves on the plot gives
the value of $T^*$. In our case the first intersection is associated with the
domain ``2''. Thus, $T^* \approx 0.06t$ for $H_z=0.03t$. Note that at the value
of the external field close to the optimal one, curves standing for domains ``1''
and ``2'' practically coincide, whereas at large values of $H_z$ the value of
$T^*$ is not governed by the domain ``2'' but ``1''.

Making use of the data presented in Fig.4 we can plot for the NOT-AND
gate the phase diagram
``critical temperature vs. stabilizing external field'', see Fig.5.
The region ``1'' corresponds to the parameter range at which the
NOT-AND gate works. The operation of the gate is not possible for the region ``2''.
As follows from Fig.5, if the value of
$H_z$ is small enough, the operation of the NOT-AND gate
at finite temperatures may be stabilized
by the increase in $H_z$. But if $H_z$ is close to its optimal
value the temperature enhancement results in the break down
of the operation of this gate.

As follows from Figs.4,5, to provide the operation of the NOT-AND
gate, the inequality $T^* < H_z$ must be fulfilled. But in reality
more stringent condition
\begin{equation}
T^* \ll H_z
\end{equation}
must be met in order to decrease the
probability to read off the erroneous result of calculations
($S_t$ must be close to unity;
$S_t = 0.5$ gives the correct result to a probability of 75 per cent only).

Let us estimate the value of $T^*$. Taking into account the theoretical
estimations \cite{singledyn} of the Coulomb repulsion between electrons on
a quantum dot $U \approx 10^{-2}$eV (that is in
agreement with the experimental data \cite{Meurer}, and the condition
for strong antiferromagnetic correlations in the system $t \ll U$ we
obtain $t \sim 10^{-3}$eV. Thus, for the realistic values of
magnetic field $H_z \sim 10^{-2}t \sim 10^{-5}$eV $\sim 1$T  the
condition (3) gives $T^* \sim 0.01 \div 0.1$K. Although such low temperatures
may be reached in experiment, it is much lower than even
the temperature of liquid helium, so the practical implementation
of computations
on the basis of quantum-dot spin gates seems to be hardly possible.
Thus, even the fundamental limitations inherent for quantum-dot gates
(such as low operation speed) or for the ground state calculation concept
itself
(the possibility for the system to be trapped in a local minima giving
rise to the erroneous result of calculation) would be circumvented,
the effect of finite temperatures is very serious obstacle for the
ground state calculations in quantum-dot spin gates.

Notice, that the approach based on non-dissipative dynamics of
interacting electrons in tunnel-coupled quantum dots \cite{Openov1},
\cite{Openov2}
may be used for implementation of various logical functions
in quantum dot spin gates in an extremely short time for
realistic magnetic fields.
Though for the non-dissipative computing concept there remain
problems to be solved (e.g. how to prepare the system in the
predetermined initial state, and to read off the result of calculation)
this approach appears to be perspective for quantum dot
systems.

To conclude, we have studied the effect of temperature on
the operation of the elementary quantum-dot spin gates for single-electron
computing within the framework of
the Hubbard model by the example of the NOT-AND gate.
The calculated values of the uniform external magnetic field necessary
to realize the whole truth table
proved to be unreasonably high for the practical implementation of
the NOT-AND logical function even at the liquid helium temperature.
Though in the present paper we list results for the NOT-AND gate
only,
for other logical gates investigated by us (e.g. AND, OR, NOR, NXOR)
the critical temperature is the same to the order of magnitude
\cite{tobe-m}.
Thus, this result appears to be common to all spin gates, so
finite temperatures seem to be a serious obstacle
for the ground state calculation concept in quantum dot spin gates.

\vskip 6mm

{\bf Acknowledgments}

This work was supported in part by the Russian Foundation for Fundamental
Research under Grant No 96-02-18918 and by the Russian State Program
``Advanced Technologies and Devices in Micro- and Nanoelectronics'' under
Grant No 02.04.329.89.5.3, and State Program ``Integratsia''. We would like to
thank L.A.Openov, S.N.Molotkov, and S.S.Nazin for fruitful discussions.

\newpage
\centerline{\bf FIGURE CAPTIONS}
\vskip 2mm

Fig.1. Three-dot NOT-AND gate. (a) Physical implementation;
(b) Logical truth table.

Fig.2. Physical truth table of the NOT-AND gate. $U/t=20, H_z/t=0.01$,
$S_t=0.95$, $T=0$. The four characteristic points (see text) are indicated.

Fig.3. Physical truth tables of the NOT-AND gate at various temperatures.
$U/t=20, H_z/t=0.05$, $S_t=0.5$. a) $T/t=0$; b) $T/t=0.08$; c) $T/t=0.085$;
d) $T/t=0.095$.

Fig.4. Temperature dependence of
modulus of $\langle\hat{S}_{z}\rangle$ on the
output dot for the characteristic points (see Fig.2) in the physical truth table
at various values of $H_z$.
Solid line corresponds to the point ``1'', dashed line stands for
the point ``2'', long dashed line for ``3''.
a) $H_z/t=0.01$; b) $H_z/t=0.03$; c) $H_z/t=0.05$; d) $H_z/t=0.07$.

Fig.5. The phase diagram ``critical temperature $T^*$ vs. stabilizing external
magnetic field $H_z$'' for the NOT-AND gate.
The region ``1'' corresponds to the parameter range at which the
NOT-AND gate works. The operation of the gate is not possible
for the region ``2''.

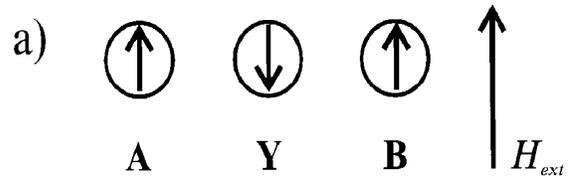
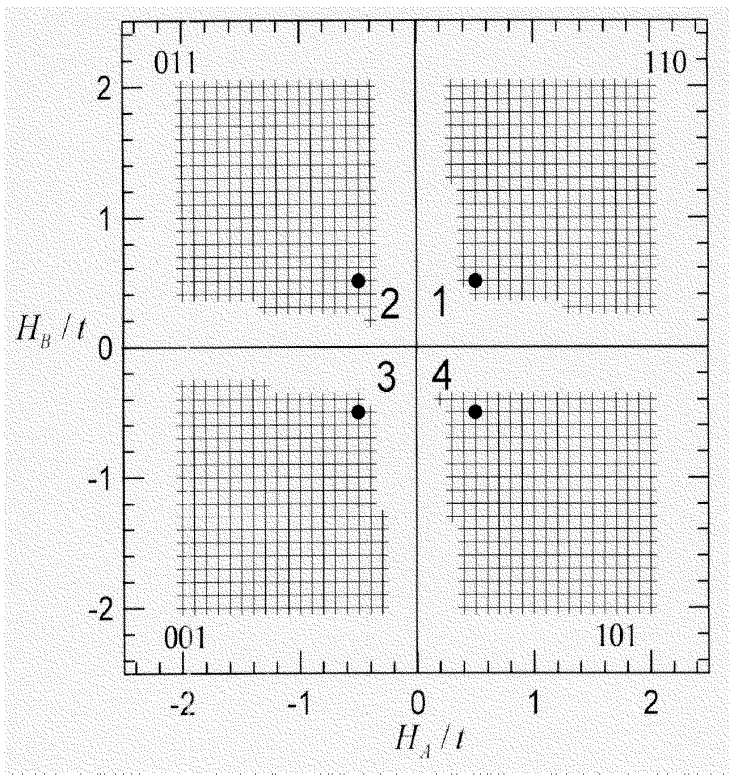
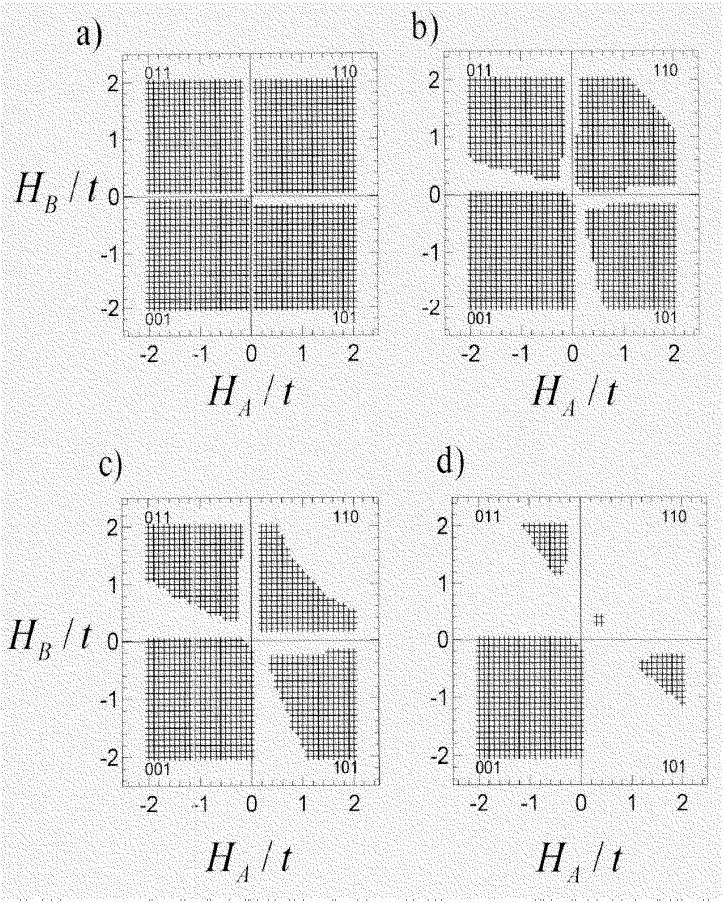
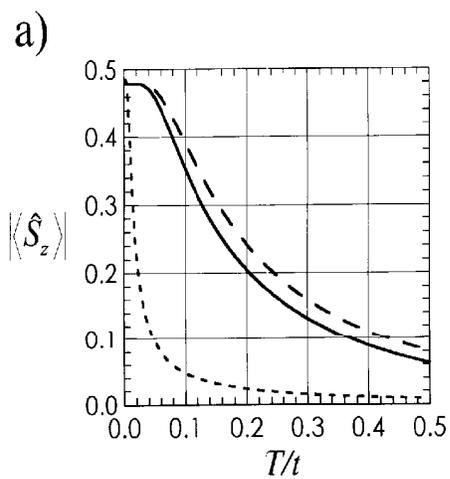
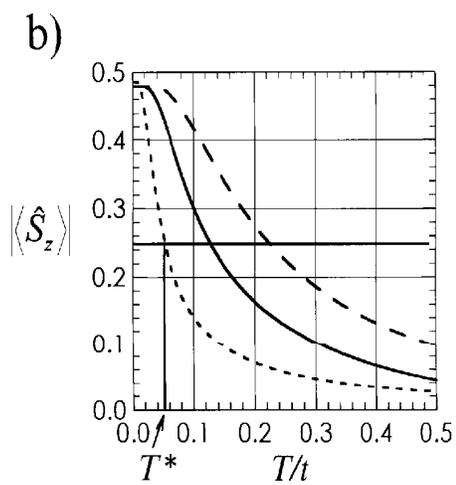
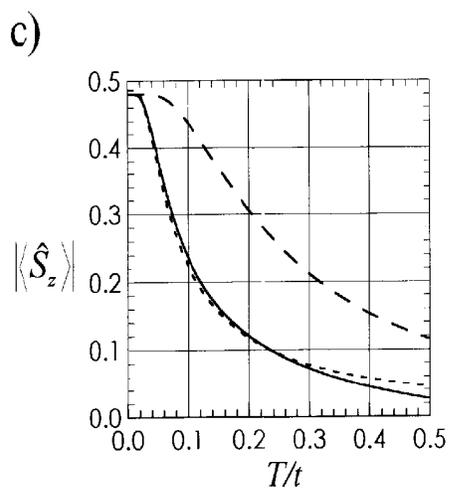
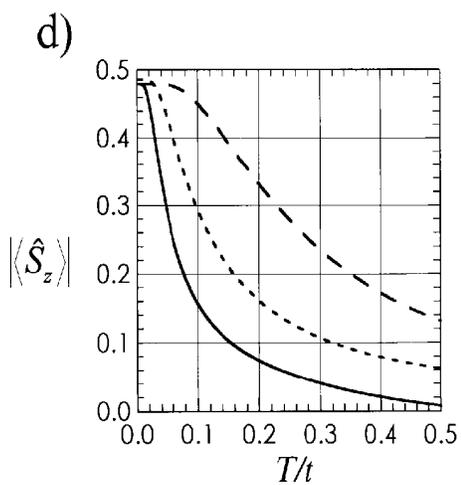
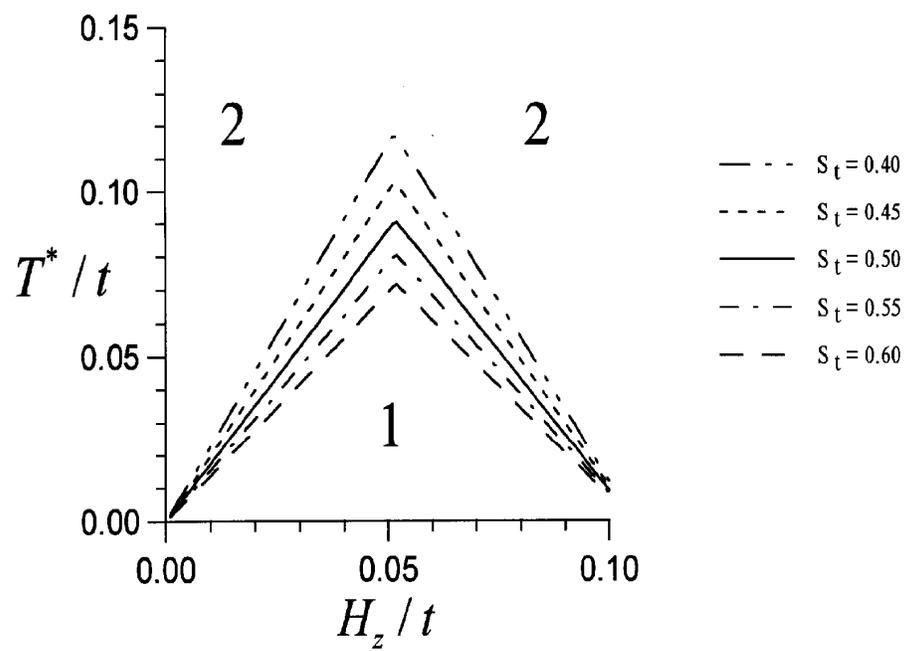
\end{document}